\documentclass[12pt,a4paper]{article}
\usepackage{a4wide}
\usepackage{amsmath}
\usepackage{amsfonts}
\usepackage{cite}
\usepackage{graphicx}
\usepackage{setspace}
\begin{document}

\title{Criteria for non-$k$-separability of $n$-partite quantum states}
\author{{N. Ananth$^a$, V. K. Chandrasekar$^b$ and  M. Senthilvelan$^a$ }\\
{\small $^a$Centre for Nonlinear Dynamics, School of Physics, Bharathidasan University,} \\{\small Tiruchirappalli - 620024, Tamilnadu, India} 
\\{\small $^b$Centre for Nonlinear Science \& Engineering, School of Electrical \& Electronics Engineering,} \\
{\small SASTRA University, Thanjavur - 613401, Tamilnadu, India}}
\date{}
\maketitle

%\section*{Abstract}
\begin{abstract}
We develop separability criteria to identify  
non-$k$-separability $(k = 2,3,\ldots ,n)$ and genuine multipartite entanglement in different classes of mixed $n$-partite quantum states 
using elements of density matrices. 
With the help of these criteria, we detect non-$k$-separability in $n$-qudit GHZ and W states respectively added with white noise. 
We also discuss the experimental implementation of our criteria by means of local observables.  
\end{abstract}

\section{Introduction}
\label{intro}
Though the $n$-number of quantum systems can have various kinds of entanglement, the focusing point is the genuine multipartite 
entanglement because it can be used for various quantum information and computational tasks \cite{horo2009,guhne2009}. An exponential 
speed-up of quantum computation requires multipartite entanglement \cite{jozsa2003}. Two widely studied multipartite entangled states 
are  (i) Greenberger-Horne- Zeilinger (GHZ) and (ii) W states. 
These two states are inequivalent and maximally entangled ones which are found 
applications in diverge topics including  quantum teleportation \cite{karl1998}, quantum secret sharing \cite{hill1999}, superdense 
coding \cite{agra2006}, splitting quantum information \cite{zheng2006} and enhancing the computational power \cite{hond2006}. The 
stronger nonlocality displayed by these two multipartite entangled states also lead to many theoretical and experimental interests in 
quantum physics, see for example Refs.\cite{green1990,banc2009,cabel2002}.

Identifying entanglement in the arbitrary multipartite states is not an easy task because 
in these systems one encounters many types of multiparticle entanglement. 
For example, the multipartite states may posses partially separable or $k$-separable 
and partially entangled or $k$-party entangled states \cite{horo2009,guhne2009}.  
An $n$-partite system is $k$-separable if it can be separated into $k$-parts. 
For example, a $4$-partite state, $ABCD$, is $3$-separable if it can be separated into any one of the following forms, 
namely $A|B|CD$, $A|C|BD$, $A|D|BC$, $B|C|AD$, $B|D|AC$ and $C|D|AB$. More precisely, an $n$-partite pure quantum state 
$|\psi_{k-\textrm{sep}}\rangle$ is called $k$- separable 
$(k=2,3,\ldots,n)$ if and only if it can be written as a product of $k$ substates, that is 
\begin{eqnarray}
\label{k0} |\psi_{k-\textrm{sep}}\rangle = |\psi_1\rangle \otimes |\psi_2\rangle \otimes \ldots \otimes|\psi_k\rangle,   
\end{eqnarray}
where $|\psi_i\rangle$, $i=1,2,\ldots,k$, represents the state of a single subsystem or a group of subsystems \cite{gabr2010}. 
A mixed state $\rho_{k-\textrm{sep}}$ is called $k$-separable, if it can be decomposed into pure $k$-separable states, that is 
\begin{eqnarray}
\label{k2} \rho_{k-\textrm{sep}} = \sum_i p_i~ \rho_{k-\textrm{sep}}^i,  
\end{eqnarray}
where $\rho_{k-\textrm{sep}}^i$ might be $k$-separable under different partitions, $p_i > 0$ and $\sum_i p_i = 1$. 
An $n$-partite state is fully separable if $k=n$ and biseparable if $k=2$.  
States that are not fully separable and not biseparable are called nonseparable and genuinely multipartite entangled states respectively. 

The aim of the present work is to identify non-$k$- separability in the GHZ and W classes of multipartite states. 
Several conditions were proposed to detect genuine multipartite entanglement and nonseparability of multipartite states    
\cite{gabr2010,dur1999,dur2000,dur2001,seev2002,uff2002,lask2005,toth2005,seev2008,hube2010,gao2010,guhne2010,gao2011,gao2013}.
To name a few, we cite the following : Seevinck and Uffink have proposed a set of inequalities, 
which can characterize various levels of partial separability and entanglement in multiqubit states \cite{seev2008}. 
Huber {\it et.al.} have proposed a general framework to obtain bilinear inequalities which can characterize 
the genuinely multipartite entangled mixed quantum states in arbitrary-dimensional systems \cite{hube2010}. 
From the later Gabriel {\it et.al.} have developed an easily computable criterion to detect $k$-nonseparability in mixed multipartite 
states \cite{gabr2010}. Recently G\"uhne and Seevinck have proposed the biseparability and full separability criteria for different 
classes of $3$-qubit and $4$-qubit states \cite{guhne2010}. These conditions were associated with density matrix elements. 
Later Gao and Hong have generalized the separability criteria proposed by G\"uhne and Seevinck to $n$-qubit and $n$-qudit 
states and proved that their criteria is applicable for any partitions \cite{gao2011}. 
The $k$-nonseparability criteria for the arbitrary dimensional mixed multipartite states 
was subsequently developed in Ref.\cite{gao2013}. 
In the present work, we extend the criteria given by Gao and Hong \cite{gao2011} to $k$-separable $n$-partite states. 
For a given $k$, violation of our criteria reveals the non-$k$-separability. 
With the help of our criteria one can detect non-$k$-separability in different classes of 
arbitrary dimensional $n$-partite states. 
We also illustrate the non-$k$-separability of mixed $n$-partite states with two examples.     
We formulate the separability conditions in terms of density matrix elements since these elements  
can be measured efficiently with local observables \cite{seev2008,gao2010,guhn2007,lu2013}.  
The conditions presented in this paper are experimentally implementable without a full quantum state tomography. 
We also discuss how many local observables are required to implement the present criteria in experiments. 
 
The paper is organized as follows. In the following section, we derive separability criteria to identify non-$k$-separable mixed $n$-partite quantum states. 
In Sec.\ref{sec3} we illustrate our criteria by considering $n$-qudit GHZ and W states respectively mixed with 
white noise. In Sec.\ref{sec4} we calculate the number of local observables required to evaluate the criteria given in this work. 
Finally, we summarize our conclusions in Sec.\ref{con}.    
%%%%%%%%%%%%%%%%%%%%%%%%%%%%%%%%%%%%%%%%%%%%%%%%%%%%%%%%%%%%%%%%%%%%%%%%%%%%%%%%%%%%%%%%%%%%%%%%%%%%%%%%%%%%%%%%%%%%%%%%%%%%%%%%%%%%%%%%%%%%%%%%%%%%%%%
\section{Criteria for non-$k$-separability }
\label{sec2}
In this section, we present the separability criteria to identify different classes of non-$k$-separable $n$-qudit states.  
We derive these conditions based on the ideas given in Refs.\cite{guhne2010,gao2011}. 
To begin we present the separability condition which is applicable for a class of GHZ multipartite states. \\ \\ 
{\bf Criterion 1. } Let $\rho$ be a $k$-separable $n$-partite density matrix acting on Hilbert space 
$\mathcal{H}_1\otimes \mathcal{H}_2 \otimes \ldots \otimes \mathcal{H}_n$, where dim $\mathcal{H}_l=d_l$, $l=1,2,\ldots,n$. Then  
\begin{align}
 \label{a5} (2^{k-1}-1)~|\rho_{1,d_{1}d_{2}...d_{n}}| \leq\frac{1}{2}\sum_{j\in A} \sqrt{\rho_{j,j}\rho_{d_{1}d_{2}...d_{n}-j+1,d_{1}d_{2}...d_{n}-j+1}}.  
\end{align}
Here $A=\{\sum_{l=1}^{n-1}j_ld_{l+1}\cdots d_n+j_n+1 ~ | ~ j_l=0, d_l-1, (j_1,j_2,\cdots, j_n)\neq
(0,0,\cdots,0),(d_1-1,d_2-1,\cdots,d_n-1)\}$. 
An $n$-partite state $\rho$ which violates the inequality (\ref{a5}), is a non-$k$-separable $n$-partite state. 
Suppose a state violates the inequality (\ref{a5}) for $k=2$, then $\rho$ is a non-$2$-separable $n$-partite state or 
a genuinely $n$-partite entangled state \cite{gao2013}.

We obtain the above inequality (\ref{a5}) from the biseparability of $n$-qudit case \cite{gao2011}. 
The inequality given above can be verified in the same manner as the Theorem 2 in Ref.\cite{gao2011} was proved. 
Since the underlying ideas are exactly the same we do not present the details here. 
The term $(2^{k-1}-1)$ which appear additionaly in (\ref{a5}) decides the non-$k$-separability
of $n$-partite states. 

In the following, we formulate another criterion applicable for a class of $n$-qudit W states \cite{kim2008}, 
which is not discussed in the earlier works \cite{guhne2010,gao2011}. 
We mention here that the $n$-qudit W class state has several generalizations and in this work   
we consider only one generalization which was considered in Ref.\cite{kim2008}.
To derive the condition for the $n$-qudit W states, we generalize the biseparability of $n$-qubit case \cite{gao2011} to $n$-qudit case 
and obtain the following form of inequality which is suitable for non-$k$-separable $n$-qudit states.  \\ \\
{\bf Criterion 2.} Let $\rho = (\rho_{i,j})_{d^n\times d^n}$ be an $n$-qudit density matrix. If $\rho$ is $k$-separable, then its density matrix elements fulfill 
\begin{align}
\label{t3w} \sum\limits_{{1\leq j<i\leq n},\atop p,q=1,2,\ldots,d-2,d-1}|\rho_{p\times d^{n-i}+1,q\times d^{n-j}+1}| 
 \leq&\sum\limits_{{1\leq j<i\leq n},\atop p,q=1,2,\ldots,d-2,d-1}\sqrt{\rho_{1,1}\rho_{p\times d^{n-i}+q\times d^{n-j}+1,p\times d^{n-i}+q\times d^{n-j}+1}} \notag\\ 
&\quad +\left(\frac{n-k}{2}\right) \sum\limits_{1\leq i\leq n,\atop p=1,2,\ldots,d-2,d-1}\rho_{p\times d^{n-i}+1,p\times d^{n-i}+1}. 
\end{align}
An $n$-qudit state $\rho$ which violates the inequality (\ref{t3w}), 
is a non-$k$-separable $n$-partite state. 
If the inequality (\ref{t3w}) is violated for $k=2$, then the state is genuinely $n$-partite entangled one. 

The criterion given in (\ref{t3w}) can also be verified in the same manner as the Theorem 3 in Ref.\cite{gao2011} was proved. 
One can deduce the non-$k$-separability criteria for $n$-qubit states by restricting $d = 2$ in Eqs.(\ref{a5}) and (\ref{t3w}).
We conclude this section by noting that in the criteria $1$ and $2$, 
$\rho_{i,j}$ represents the $i^{\text{th}}$ row and $j^{\text{th}}$ 
column element in the density matrix.
%%%%%%%%%%%%%%%%%%%%%%%%%%%%%%%%%%%%%%%%%%%%%%%%%%%%%%%%%%%%%%%%%%%%%%%%%%%%%%%%%%%%%%%%%%%%%%%%%%%%%%%%%%%%%%%%%%%%%%%%%%%%%%%%%%%%%%%%%%%%%%%%%%%%%%%%%%%%%%%%%%%%
\section{Examples}
\label{sec3}
In this section, we analyze the non-$k$-separability of $n$-partite GHZ state mixed with white noise through the criterion 1.  
We then investigate the non-$k$-separability of $n$-partite W state mixed with white noise through the criterion 2. 
In both the examples we consider the $3$-qutrit and $4$-qutrit cases and explain their nonseparability and genuine multipartite entanglement in detail. 
%%%%%%%%%%%%%%%%%%%%%%%%%%%%%%%%%%%%%%%%%%%%%%%%%%%%%%%%%%%%%%%%%%%%%%%%%%%%%%%%%%%%%%%%%%%%%%%%%%%%%%%%%%%%%%%%%%%%%%%%%%%%%%%%%%%%%%%%%%%%%%%%%%%%%%%%%%%%%%%%%%%%%%
\subsection{$n$-qudit GHZ state mixed with white noise}
\begin{figure}[t]
\begin{center}
\includegraphics[width=0.5\linewidth]{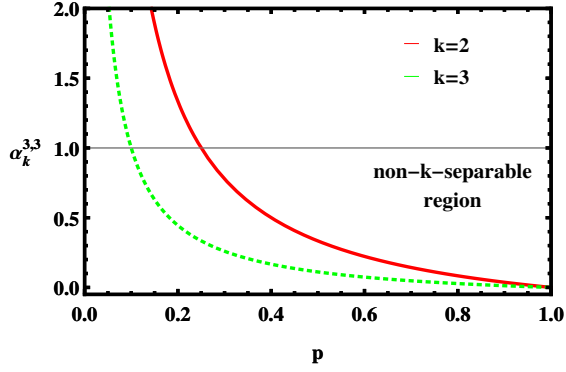}
\caption{non-$k$-separability of $3$-qutrit GHZ state mixed with white noise} \label{f1}
\end{center}
\end{figure}

\begin{figure}[t]
\begin{center}
\includegraphics[width=0.5\linewidth]{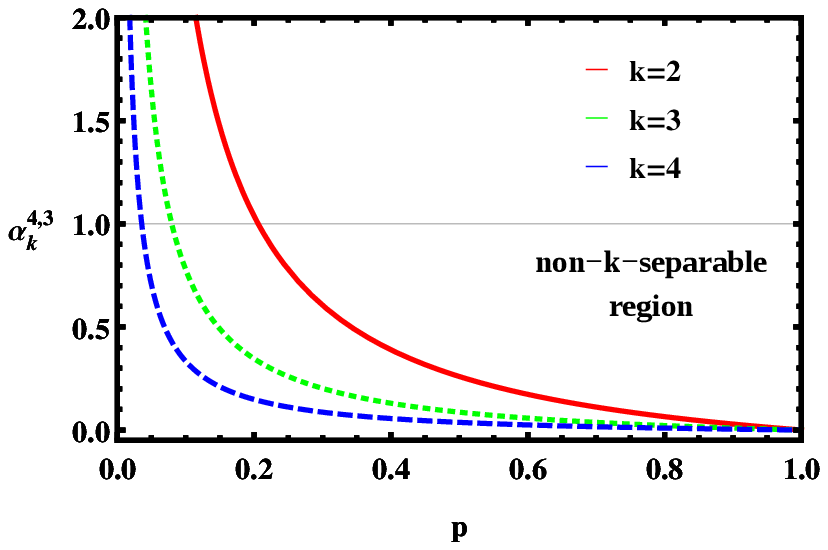}
\caption{non-$k$-separability of $4$-qutrit GHZ state mixed with white noise} \label{f2}
\end{center}
\end{figure}

To illustrate the criterion 1, we consider the $n$-qudit GHZ state mixed with white noise, 
\begin{align}
\label{rdn}\rho_{dn} = p|GHZ_{dn}\rangle\langle GHZ_{dn}| + \frac{(1-p)} {d^n} I, 
\end{align}
where $|GHZ_{dn}\rangle $ $= \frac{1}{\sqrt{d}}\sum_{i=0}^{d-1}|i\rangle^{\otimes n}$ and 
$I$ is the Identity operator \cite{hube2010}. 
Imposing the condition (\ref{a5}) on the state (\ref{rdn}), we can obtain the following general function, namely    
\begin{align}
\label{alph}\alpha_k^{n,d} = \frac{2^{n-1}-1}{2^{k-1}-1}\times \frac{1-p}{p\times d^{n-1}}.
\end{align}
The outcome $\alpha_k^{n,d} < 1$, for the given value of $k$ $(k=2,3,\ldots,n)$,   
confirms that the state is non-$k$-separable. 
To illustrate the non-$k$-separability, let us consider the $3$-qutrit ($n=3$ and $d=3$) and 
$4$-qutrit ($n=4$ and $d=3$) cases in (\ref{rdn}). For these two cases, Eq.(\ref{alph}) turns out to be 
$\alpha_k^{3,3}=\frac{(1-p)}{3p(2^{(k-1)}-1)}$ and $\alpha_k^{4,3}=\frac{7(1-p)}{27p(2^{(k-1)}-1)}$.  
We plot these two functions for various $k$ $(2\leq k\leq n)$ values and depict the outcome in Figs.\ref{f1} and \ref{f2} \cite{hube2013}.  
In these two Figures, the region covered by $\alpha_k^{n,d} < 1$ brings out the non-$k$-separability.   
For the state (\ref{rdn}), the criterion (\ref{a5}) act as strong as the PPT criterion and the criteria developed in Refs.\cite{gabr2010,gao2010}
for detecting nonseparable quantum states. 
\subsection{$n$-qudit W state mixed with white noise} 
To illustrate the criterion 2, we consider the 
$n$-qudit W state with additional isotropic (white) noise, 
\begin{align} 
\label{rwn}\rho_{W_n} = (1-p) |W_n^d\rangle \langle W_n^d| + p \frac{I}{d^n}, 
\end{align}
where $|W_n^d\rangle = \frac{1}{\sqrt{n\times (d-1)}} \big({\sum_{i=1}^{d-1}}$ $(|00\ldots i\rangle + |0\ldots i0\rangle$ $+ \cdots 
+|i0\ldots 0\rangle)\big)$ and $I$ is the identity operator. 
Applying the inequality (\ref{t3w}) on the state $\rho_{W_n}$, given above, we find 

\begin{align}
\label{bet} \beta_k^{n,d} = \left(\frac{p~n~(d-1)}{d^n~(1-p)}\right)+ \left(n(d-1)+\frac{n^2(d-1)^2~p}{d^n~(1-p)}\right) 
\times \left(\frac{n-k}{2}\right)\times\frac{1}{\left(\sum\limits_{i=1}^{n(d-1)-1} i - n \sum\limits_{j=1}^{d-2} j\right)}. 
\end{align}
An $n$-partite state (\ref{rwn}) is non-$k$-separable if it obeys the inequality $\beta_k^{n,d} < 1$ for a given $k$. 
In other words the genuine multipartite entanglement of $\rho_{W_n}$ can be confirmed with $\beta_2^{n,d} < 1$.
\begin{figure}[t]
\begin{center}
\includegraphics[width=0.5\linewidth]{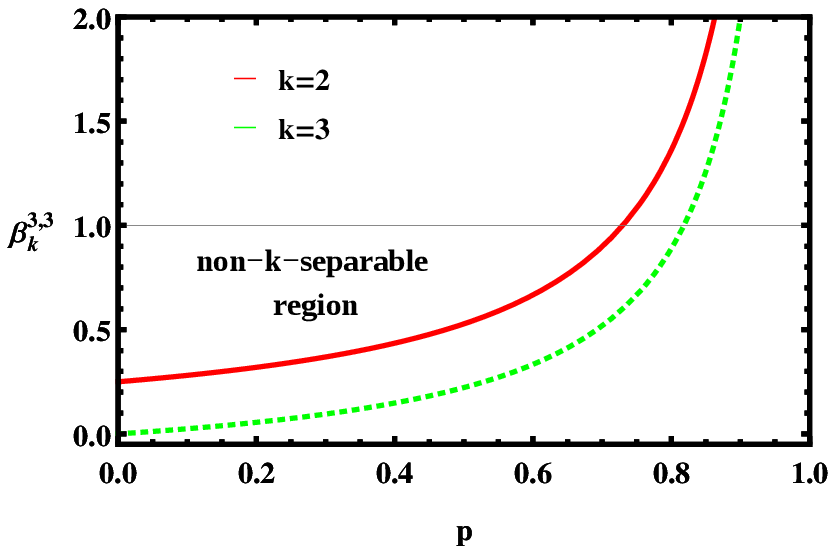}
\caption{non-$k$-separability of $3$-qutrit W state mixed with white noise}  \label{f3}
\end{center}
\end{figure}
\begin{figure}[t]
\begin{center}
\includegraphics[width=0.5\linewidth]{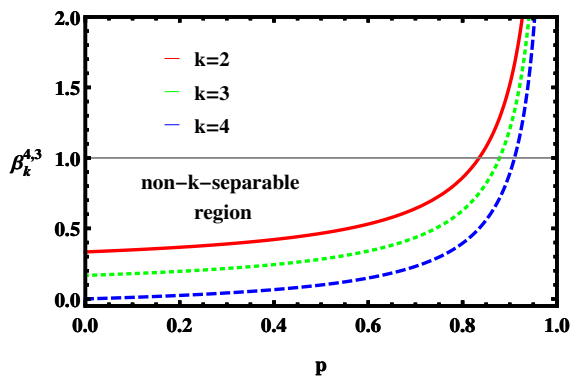}
\caption{non-$k$-separability of $4$-qutrit W state mixed with white noise}  \label{f4}
\end{center}
\end{figure}
To identify the non-$k$-separabilities of $3$-qutrit and $4$-qutrit mixed states $\rho_{W_n}$ respectively in (\ref{rwn}), 
we consider the functions $\beta_k^{3,3}$ ($n=3,d=3$ in Eq.(\ref{bet})) and $\beta_k^{4,3}$ ($n=4,d=3$ in Eq.(\ref{bet})). 
We analyze these two functions for various $k$ values and plot the results in Figs.\ref{f3} and \ref{f4}.  
%%%%%%%%%%%%%%%%%%%%%%%%%%%%%%%%%%%%%%%%%%%%%%%%%%%%%%%%%%%%%%%%%%%%%%%%%%%%%%%%%%%%%%%%%%%%%%%%%%%%%%%%%%%%%%%%%%%%%%%%%%%%%%%%%%%%%%%%%%%%%%%%%%%%%%% 
\section{Experimental feasibility}
\label{sec4} 
We formulated the separability criteria in terms of density matrix elements. 
The condition given above can also be experimentally accessible by means of local observables such as 
$\mathcal{L}=\mathcal{A}_1\otimes \mathcal{A}_2 \otimes\ldots\otimes \mathcal{A}_n$, 
where $\mathcal{A}_l$ acts on $l^{\text{th}}$ subsystem. 
In the following, we calculate the required number of local observables to implement the matrix elements,  
which are present in the inequalities (\ref{a5}) and (\ref{t3w}), in experiments.  
For this purpose, we redefine the observables given in Ref.\cite{gao2010} to 
determine the elements in higher dimensional multipartite states.  

Following the method given in Refs.\cite{seev2008,gao2010,guhn2007}, we determine the modulus of the far-off antidiagonal element, 
$|\rho_{1,d_1d_2\ldots d_n}|$, present in Eq.(\ref{a5}), by measuring the observables $Q$ and $\tilde{Q}$.  
The operators $\langle Q\rangle = 2 \text{Re} (\rho_{1,d_1d_2\ldots d_n})$ and  
$\langle\tilde{Q}\rangle = -2 \text{Im} (\rho_{1,d_1d_2\ldots d_n})$ can be represented as 
\begin{subequations}
\begin{align}
Q=& {|0\rangle\langle (d_1-1)(d_2-1)\ldots (d_n-1)|}^{\otimes n} + {|(d_1-1)(d_2-1)\ldots (d_n-1)\rangle\langle 0|}^{\otimes n}, \\  
\tilde{Q}=& -i{|0\rangle\langle (d_1-1)(d_2-1)\ldots (d_n-1)|}^{\otimes n} + i {|(d_1-1)(d_2-1)\ldots (d_n-1)\rangle\langle 0|}^{\otimes n}. 
\end{align}
\end{subequations}
Then the far-off antidiagonal element can be obtained from two measurement settings $\mathcal{M}_l$ and $\tilde{\mathcal{M}_l}$, given by 
\begin{subequations}
\begin{align}
\label{ml} \mathcal{M}_l =& \bigotimes_{j=1}^n \left[\cos\left(\frac{l\pi}{n}\right) R_l^j + \sin\left(\frac{l\pi}{n}\right) \tilde{R}_l^j\right],
\\
\label{mlt} \tilde{\mathcal{M}_l} =& \bigotimes_{j=1}^n \left[\cos\left(\frac{l\pi+\frac{\pi}{2}}{n}\right) R_l^j + 
\sin\left(\frac{l\pi+\frac{\pi}{2}}{n}\right) \tilde{R}_l^j\right],
\end{align}
\end{subequations}
where $R_l^j = |y_l^j\rangle\langle x_l| + |x_l\rangle\langle y_l^j|$, 
$\tilde{R}_l^j = i |y_l^j\rangle\langle x_l| - i |x_l\rangle\langle y_l^j|$, 
$|x_l\rangle = |0 \rangle$, $|y_l^j\rangle = |d_j-1\rangle$, $d_j$ is the dimension of the $j^{\text{th}}$ subsystem,  
$j=1,2,\ldots,n$ and $l=1,2,\ldots,n$. The operators (\ref{ml}) and (\ref{mlt}) also obey  
\begin{align}
\sum_{l=1}^n (-1)^l \mathcal{M}_l = n Q, \qquad
\sum_{l=1}^n (-1)^l \tilde{\mathcal{M}_l} = n \tilde{Q}, 
\end{align}
which can be verified in the same way as done in Ref.\cite{guhn2007}. 
Therefore, the real and imaginary parts of an antidiagonal 
element of $n$-partite state can be determined by $2n$ local observables. 

Now we determine the modulus of the off-diagonal elements, $|\rho_{p d^{n-i}+1,q d^{n-j}+1}|$, which appear in the left hand side of 
inequality (\ref{t3w}) by measuring the observables $O_{ab}^{rs}$ and $\tilde{O}_{ab}^{rs}$, where 
$\langle O_{ab}^{rs}\rangle = 2\text{Re}(\rho_{p d^{n-i}+1,q d^{n-j}+1})$ and 
$\langle \tilde{O}_{ab}^{rs}\rangle = -2\text{Im}$ $(\rho_{p d^{n-i}+1,q d^{n-j}+1})$. 
Without loss of generality, let $r<s$, they can be written as  
\begin{subequations}
\begin{align}
O_{ab}^{rs} =& \frac{1}{2} T^{\otimes (r-1)} \otimes M_a \otimes T^{\otimes (s-r-1)} \otimes N_b \otimes T^{\otimes (n-s)}\qquad\qquad \notag\\
 &+\frac{1}{2} T^{\otimes (r-1)} \otimes \tilde{M}_a \otimes T^{\otimes (s-r-1)} \otimes \tilde{N}_b \otimes T^{\otimes (n-s)},\\ 
\tilde{O}_{ab}^{rs} =& \frac{1}{2} T^{\otimes (r-1)} \otimes M_a \otimes T^{\otimes (s-r-1)} \otimes \tilde{N}_b \otimes T^{\otimes (n-s)} \notag\\
 &-\frac{1}{2} T^{\otimes (r-1)} \otimes \tilde{M}_a \otimes T^{\otimes (s-r-1)} \otimes N_b \otimes T^{\otimes (n-s)}. 
\end{align}
\end{subequations}
Here $T=|x\rangle\langle x|$, $M_a = |a\rangle\langle x| + |x\rangle\langle a|$, $\tilde{M}_a = i|a\rangle\langle x| - i|x\rangle\langle a|$, 
$N_b = |b\rangle\langle x| + |x\rangle\langle b|$, $\tilde{N}_b = i |b\rangle\langle x| - i |x\rangle\langle b|$, $x=0$ and  
$a,b = \{ 1,2,\ldots,d-1\}$. Therefore, the off-diagonal element can be determined by measuring 
the real and imaginary parts in which each one is associated with two local observables. 
Therefore, the term which appear in the left hand side of the inequality (\ref{t3w}) can be determined 
by $4(d-1)\sum_{i=1}^{n-1} i(d-1)$ local observables. 

Finally, the diagonal elements that present in the right hand side of expressions (\ref{a5}) and (\ref{t3w}) 
can be implemented by the following local observables, namely 
\begin{align}
\label{diag} |x_1 x_2 \ldots x_n\rangle\langle x_1 x_2 \ldots x_n | = \bigotimes_{i=1}^n T_{m_i}, 
\end{align}
with $T_{m_i}= |m_i\rangle\langle m_i |$, $m_i = 0,1,2,\ldots,d_i-1$. 
It is clear from (\ref{diag}) that to determine a diagonal matrix element, it is required only one local observable. 

We note here that for the criterion 1 eventhough the total number of density matrix elements of an $n$-partite state is 
$d_1^2\times d_2^2\times d_3^2\times\ldots\times d_n^2$, 
we need to measure only $2^n-1$ elements out of it.  
They require $2^n+2n-2$ local observables in order to identify the non-$k$-separability by the criterion $1$.  
Similarly for the criterion 2 eventhough the total number of density matrix elements of an $n$-partite state is $d^{2n}$,   
we need to measure only $2\times(d-1)\sum_{i=1}^{n-1} (i\times (d-1)) + (n\times(d-1)+1)$ elements out of it. 
In other words one totally requires $5(d-1)\sum_{i=1}^{n-1} i(d-1)+(n(d-1)+1)$ local observables to test the criterion $2$.  
Since the elements need to be measured are very less compared to the total number of elements, 
it would require only fewer measurements compared to the $(d_1^2-1)(d_2^2-1)\ldots(d_n^2-1)$ number of measurements needed for quantum state tomography. 
%%%%%%%%%%%%%%%%%%%%%%%%%%%%%%%%%%%%%%%%%%%%%%%%%%%%%%%%%%%%%%%%%%%%%%%%%%%%%%%%%%%%%%%%%%%%%%%%%%%%%%%%%%%%%%%%%%%%%%%%%%%%%%%%%%%%%%%%%%%%%%%%%%%%%%%
\section{Conclusion}
\label{con}
In this work, we have extended the criteria given by Gao and Hong to $k$-separable $n$-partite states. 
With the help of our criteria $1$ and $2$ one can identify the non-$k$-separability $(k = 2,3,\ldots ,n)$ 
and genuine $n$-partite entanglement in mixed quantum states. 
We have verified non-$k$-separability of different classes of mixed multipartite states. 
We have also given two general functions namely, $\alpha_k^{n,d}$ and $\beta_k^{n,d}$, 
to detect non-$k$-separability in the $n$-qudit GHZ state and  
$n$-qudit W state, respectively added with white noise.  
Our criteria can also identify the nonseparability of mixture of GHZ and W states added with white noise. 
We have also shown that the criteria developed in this paper can be computable and implementable in experiments. 
They require only fewer measurements compared to full quantum state tomography. \\

%%%%%%%%%%%%%%%%%%%%%%%%%%%%%%%%%%%%%%%%%%%%%%%%%%%%%%%%%%%%%%%%%%%%%%%%%%%%%%%%%%%%%%%%%%%%%%%%%%%%%%%%%%%%%%%%%%%%%%%%%%%%%%%%%%%%%%%%%%%%%%%%%%%%%%%
\noindent Acknowledgement : We would like to thank the referee for his valuable suggestions to improve the quality of this paper. 
%%%%%%%%%%%%%%%%%%%%%%%%%%%%%%%%%%%%%%%%%%%%%%%%%%%%%%%%%%%%%%%%%%%%%%%%%%%%%%%%%%%%%%%%%%%%%%%%%%%%%%%%%%%%%%%%%%%%%%%%%%%%%%%%%%%%%%%%%%%%%%%%%%%%%%%
% Non-BibTeX users please use


\begin{thebibliography}{}
\bibitem{horo2009}
R. Horodecki, P. Horodecki, M. Horodecki and K. Horodecki, Rev. Mod. Phys. {\bf 81}, 865-942 (2009) 
\bibitem{guhne2009}
O. G\"uhne and G. T\'oth, Phys. Rep. {\bf 474}, 1-75 (2009)

\bibitem{jozsa2003}
R. Jozsa and N. Linden, Proc. R. Soc. A {\bf 459}, 2011 (2003)

\bibitem{karl1998}
A. Karlsson and M. Bourennane,  Phys. Rev. A \textbf{58}, 4394 (1998)

\bibitem{hill1999}
M. Hillery, V. Bu\v{z}ek and A. Berthiaume, Phys. Rev. A  \textbf{59}, 1829 (1999)

\bibitem{agra2006}
P. Agrawal and A. Pati, Phys. Rev. A \textbf{74}, 062320 (2006)

\bibitem{zheng2006}
S-B Zheng, Phys. Rev. A {\bf 74}, 054303 (2006)

\bibitem{hond2006}
E. D'Hondt and P. Panangaden, Quantum Inf. Comput. {\bf 6}, 173 (2006). 

\bibitem{green1990}
D. M. Greenberger, M. A. Horne, A. Shimony and A. Zeilinger, Am. J. Phys. \textbf{58}, 1131 (1990)

\bibitem{banc2009}
J. D. Bancal, C. Branciard, N. Gisin and S. Pironio, Phys. Rev. Lett. \textbf{103}, 090503 (2009)

\bibitem{cabel2002}
A. Cabello, Phys. Rev. A \textbf{65}, 032108 (2002) 

\bibitem{gabr2010}
A. Gabriel, B. C. Hiesmeyr and M. Huber, Quantum Inf. Comput. \textbf{10} 0829 (2010) 

\bibitem{dur1999}
W. D\"ur, J. I. Cirac and R. Tarrach, Phys. Rev. Lett. \textbf{83}, 3562 (1999)

\bibitem{dur2000}
W. D\"ur and J. I. Cirac, Phys. Rev. A \textbf{61}, 042314 (2000)

\bibitem{dur2001}
W. D\"ur and J. I. Cirac, J. Phys. A \textbf{34}, 6837 (2001)

\bibitem{seev2002}
M. Seevinck and G. Svetlichny, Phys. Rev. Lett. \textbf{89}, 060401 (2002)

\bibitem{uff2002}
J. Uffink, Phys. Rev. Lett. \textbf{88}, 230406 (2002)

\bibitem{lask2005}
W. Laskowski and M. Zukowski, Phys. Rev. A \textbf{72}, 062112 (2005)

\bibitem{toth2005}
G. T\'oth and O. G\"uhne, Phys. Rev. A {\bf 72}, 022340 (2005)

\bibitem{seev2008}
M. Seevinck and J. Uffink, Phys. Rev. A \textbf{78}, 032101 (2008) 

\bibitem{hube2010}
M. Huber, F. Mintert, A. Gabriel and B. C. Hiesmeyr, Phys. Rev. Lett. \textbf{104} 210501 (2010) 

\bibitem{gao2010}
T. Gao and Y. Hong, Phys. Rev. A \textbf{82}, 062113 (2010)

\bibitem{guhne2010}
O. G\"uhne and M. Seevinck, New J. Phys. \textbf{12}, 053002 (2010)
\bibitem{gao2011}
T. Gao and Y. Hong, Eur. Phys. J. D \textbf{61}, 765 (2011)

\bibitem{gao2013}
T. Gao, Y. Hong, Y. Lu and F. Yan, Europhys. Lett. \textbf{104}, 20007 (2013)

\bibitem{guhn2007}
O. G\"uhne, C-Y Lu, W-B Gao and J-W Pan, Phys. Rev. A \textbf{76}, 030305(R) (2007)

\bibitem{lu2013}
Y. Lu, G. L. Long and T. Gao, Int. J. Theor. Phys. \textbf{52}, 699 (2013)

\bibitem{kim2008}
J. S. Kim and B. C. Sanders, J. Phys. A \textbf{41}, 495301 (2008)

\bibitem{hube2013}
M. Huber, M. P. Llobet and J. I. de Vicente, Phys. Rev. A \textbf{88}, 042328 (2013)


\end{thebibliography}
\end{document}